\documentclass[
    reprint, 
    superscriptaddress, 
    amsmath, 
    amssymb, 
    aps, 
    prx
]{revtex4-2}

\usepackage{graphicx}
\usepackage{dcolumn}
\usepackage{bm}
\usepackage{hyperref}

\newcommand{\ii}{\mathrm{i}}
\newcommand{\dd}{\mathrm{d}}

\newcommand{\lr}[1]{\!\left(#1\right)\!}

\makeatletter
\newcommand*{\rom}[1]{\expandafter\@slowromancap\romannumeral #1@}

\DeclareMathOperator{\Tr}{Tr}
\DeclareMathOperator{\geff}{g_{\text{eff}}}

\usepackage{amsfonts}
\usepackage{amssymb,amstext,amsthm,eucal,amsmath}
\usepackage{mathrsfs}
\usepackage{mathtools}
\usepackage{graphicx, caption}
\usepackage{subcaption}
\usepackage{wrapfig}
\usepackage{array}
\usepackage{float}
\usepackage{enumerate}
\usepackage{upgreek}
\usepackage{listings}
\usepackage{natbib}
\usepackage{physics}
\usepackage[dvipsnames]{xcolor}
\usepackage{todonotes}
\usepackage{ragged2e}
\newcommand{\mycomment}[1]{}
\usepackage{comment}

\usepackage{tikz}
\usetikzlibrary{arrows, positioning, fadings, patterns, patterns.meta}
\usepackage{pgfplots, pgfplotstable} 
\usepgfplotslibrary{fillbetween}
\pgfplotsset{compat=1.7}

\usepackage{verbatim}
\usepackage[version=4]{mhchem} 
\usepackage{wrapfig} 
\usepackage{mwe} 
\usepackage{caption} 
\usepackage{adjustbox}
\usepackage{todonotes}

\usepackage{url}
\usepackage{bm}
\usepackage{enumitem}

\RequirePackage[labelsep=space,tableposition=top]{caption}

\usepackage{subcaption}

\usepackage{booktabs} 

\usepackage{amsfonts}
\usepackage{amsmath}
\usepackage{amssymb}
\usepackage{siunitx} 
\usepackage{braket} 
\usepackage{algorithm2e}

\usetikzlibrary{spy}

\usepackage{hyperref}

\definecolor{MatlabBlue}{rgb}{0.00000,0.44700,0.74100}
\definecolor{MatlabMagenta}{rgb}{1.00000,0.00000,1.00000}
\definecolor{IndicatorCol}{RGB}{217,94,37}
\definecolor{ChipGray}{RGB}{125,125,125}
\definecolor{ChipGold}{RGB}{240,222,54}
\definecolor{LaserRed}{RGB}{255,131,127}
\definecolor{Ggreen}{RGB}{139,195,74}
\definecolor{Gorange}{RGB}{233,189,99}
\definecolor{Gpurple}{RGB}{93,93,184}
\definecolor{Gpink}{RGB}{224,152,180}
\definecolor{LaserBlue}{RGB}{30,110,143}

\definecolor{SkyBlue}{rgb}{0.5294117647058824, 0.807843137254902, 0.9215686274509803}
\definecolor{Gblue}{RGB}{33,82,135}
\definecolor{RoyalBlue}{rgb}{0.2549019607843137, 0.4117647058823529, 0.8823529411764706}
\definecolor{Gred}{RGB}{151,43,80}
\definecolor{RoyalPink}{RGB}{243,119,131}
\definecolor{RoyalOrange}{RGB}{255,220,77}
\definecolor{Black}{RGB}{0, 0, 0}

\tikzfading[name=fade out,
inner color=transparent!0,
outer color=transparent!100]

\pgfplotsset{
  /pgfplots/colormap={pink}{%
    color(0cm) = (purple);
    color(1cm) = (pink!80!purple);
    color(2cm) = (pink!90);
    color(3cm) = (pink) }
}

\usepgfplotslibrary{groupplots}

\newcommand{\drawgrid}[8]{%
    \begin{scope}[shift={(#1,#2)}]
    \def\overlapShift{0.15pt}
    \foreach \row/\col in {#6} {
        \pgfmathsetmacro\xstart{\col-1}
        \pgfmathsetmacro\ystart{\row-1}
        \pgfmathsetmacro\xend{\col}
        \pgfmathsetmacro\yend{\row}
        \fill[#7] 
            (\xstart*#3 - \overlapShift,\ystart*#3 - \overlapShift) rectangle (\xend*#3 + \overlapShift,\yend*#3 + \overlapShift);
    }

    \ifthenelse{\equal{#8}{0mm}}{}{
        \pgfmathsetmacro{\halflinewidth}{#8/2}
        \pgfmathsetmacro{\halflinethickness}{0.01mm/2}
        
        \foreach \i in {0,...,#4} {
            \draw[line width=#8] (\i*#3, -\halflinethickness) -- (\i*#3, #5*#3+\halflinewidth);
        }
        \foreach \j in {0,...,#5} {
            \draw[line width=#8] (0, \j*#3) -- (#4*#3+\halflinewidth, \j*#3);
        }
    }
    \end{scope}
}

\def\gridLineWidth{0mm}
\def\smallSquareSize{0.08cm}

\definecolor{pcolourA}{RGB}{19, 61, 105}  
\definecolor{pcolourD}{RGB}{242, 102, 102}  
\definecolor{pcolourP}{RGB}{0, 0, 0}  

\usepackage{lipsum}   

\newcommand{\MTJ}[1]{\textcolor{orange!90}{ #1}}

\definecolor{pcolourA}{RGB}{19, 61, 105}  
\definecolor{pcolourD}{RGB}{242, 102, 102}  

\hypersetup{  colorlinks,
              linktoc         = page, 
              urlcolor        = {green!50!blue},
              linkcolor       = {Bittersweet!80!black}, 
              urlcolor        = {Bittersweet!80!black}, 
              citecolor       = {cyan!50!black}, 
              anchorcolor     = {yellow}
}

\begin{document}
\preprint{APS/123-QED}

\title{\textbf{Information in quantum field theory simulators: Thin-film superfluid helium}}

\author{Maciej~T.~Jarema}
    \email{ppymj11@nottingham.ac.uk}
    \affiliation{School of Mathematical Sciences, University of Nottingham, University Park, Nottingham, NG7 2RD, UK}
    \affiliation{Centre for the Mathematics and Theoretical Physics of Quantum Non-Equilibrium Systems, University of Nottingham, Nottingham, NG7 2RD, UK}
\author{Cameron~R.~D.~Bunney}
    \affiliation{Centre for the Mathematics and Theoretical Physics of Quantum Non-Equilibrium Systems, University of Nottingham, Nottingham, NG7 2RD, UK}
   \affiliation{School of Physics and Astronomy, University of Nottingham, University Park, Nottingham, NG7 2RD, UK}
   \affiliation{Nottingham Centre of Gravity, University of Nottingham, Nottingham, NG7 2RD, UK}
\author{Vitor~S.~Barroso}
   \affiliation{School of Mathematical Sciences, University of Nottingham, University Park, Nottingham, NG7 2RD, UK}
   \affiliation{Centre for the Mathematics and Theoretical Physics of Quantum Non-Equilibrium Systems, University of Nottingham, Nottingham, NG7 2RD, UK}
\author{Mohammadamin~Tajik}
  \affiliation{Vienna Center for Quantum Science and Technology (VCQ), Atominstitut, TU Wien, Vienna, Austria}
\author{Chris~Goodwin}
   \affiliation{School of Mathematical Sciences, University of Nottingham, University Park, Nottingham, NG7 2RD, UK}
   \affiliation{Centre for the Mathematics and Theoretical Physics of Quantum Non-Equilibrium Systems, University of Nottingham, Nottingham, NG7 2RD, UK}
\author{Silke~Weinfurtner}
   \email{silke.weinfurtner@nottingham.ac.uk}
   \affiliation{School of Mathematical Sciences, University of Nottingham, University Park, Nottingham, NG7 2RD, UK}
   \affiliation{Centre for the Mathematics and Theoretical Physics of Quantum Non-Equilibrium Systems, University of Nottingham, Nottingham, NG7 2RD, UK}
   \affiliation{Nottingham Centre of Gravity, University of Nottingham, Nottingham, NG7 2RD, UK}

\date{\today}

\begin{abstract}
    Understanding quantum correlations through information-theoretic measures is fundamental to developments in quantum field theory, quantum information, and quantum many-body physics. A central feature in a plethora of systems is the \textit{area law}, under which information scales with the size of the boundary of the system, rather than volume. Whilst many systems and regimes exhibiting an area law have been identified theoretically, experimental verification remains limited, particularly in continuous systems. We present a methodology for measuring mutual information in an experimental simulator of non-interacting quantum fields, and propose using the analogue $(2+1)$-dimensional spacetime offered by thin films of superfluid helium. We provide numerical predictions incorporating the natural thermal state of the helium sample that exemplify an area-law scaling of mutual information, and characterise deviations attributable to the inherent finite system size.
    \end{abstract}

\keywords{Quantum Information, Area Law, Mutual Information, Analogue systems, Simulators, Quantum Simulators, Superfluid Helium, Robin Boundary Conditions, Quantum Field Theory, thin film, thin film superfluid helium}

\maketitle

\section{\label{section::Introduction} Introduction}
\vspace{-4mm}

Classifying information measures between particles, spatial regions and system-environment pairs is central to many advancements in modern physics~\cite{Cover_thomas_2006, Kantz_Schreiber_2003, Wilde_2013}. Leading to numerous findings such as thermalisation of isolated systems \cite{Kaufman_2016, Islam_2015}, heat cost to computation \cite{Landauer_1961,Berut_2012_Landauer, Reeb_2014_Landauer, Aimet_2024_Landauer}, and Lieb-Robinson bounds for the maximal speed of information transfer \cite{Lieb1972, Bravyi_2006, Chessa_2019, Cheneau_2022, Tran_2020, Epstein_2017}, these measures allow for a wide, general study of system correlations.
Furthermore, information becomes a central object of focus when studying quantum communication \cite{Wilde_2013} and the structure of correlations in Quantum Field Theories (QFTs) \cite{casini2023_Et_in_QFT}. All highly motivated by the immense drive towards preparing, measuring and harvesting the power of a strong type of correlations --- quantum entanglement \cite{Horodecki_2009}.

From evaporating black holes to various quantum many-body systems, entropy scales with the bounding area of the system, not its volume --- a behaviour known as an \textit{area law}. Even within theoretical frameworks of quantum information, entropy displays an area law in many scenarios~\cite{Hawking_1974, Srednicki_1993}. There exists, however, a second quantifier, mutual information (MI) --- a measure of classical and quantum correlations --- that has been proven under specific conditions to display an area-law scaling~\cite{Wolf2008}.

Experimental studies of information measures in many-body quantum systems have, however, been limited due to the difficulty of full-state reconstruction~\cite{Cramer2010}. Using the full quantum state has been successfully implemented to quantify the von Neumann and Rényi-$2$ entropies in small-lattice quantum systems~\cite{Kaufman_2016, Islam_2015, Brydges_2019}. However, the same techniques are not applicable to continuous systems, though novel methods have been developed for Gaussian states --- those fully characterised by their covariance. This enabled verification of area-law scaling of MI in a one-dimensional thermal field-theory simulator~\cite{Tajik_MI_2023} through a tomographic read-out technique~\cite{Gluza2020}.

Advances in two-dimensional quantum systems have opened avenues for the development of $(2+1)$-dimensional quantum field theory simulators~\cite{patrick2024,Bunney2023_published, Viermann2022, Navon_2021, Spence_2021}. Precise spatially and temporally resolved measurement schemes for quantum liquid interfaces~\cite{Barroso2023,Barroso2025} such as thin-film superfluid helium~\cite{Atkins1959, Bunney2023_published} offer the possibility for investigating quantum effects on a broad range of length scales, from nanometres to centimetres. Currently, the exact boundary conditions in thin-film superfluid helium systems are less-well understood~\cite{Baker2016}, but are expected to impact upon measures of mutual information. Hence, the intricate dependence of finite-size effects is essential, and in the absence of boundary-condition knowledge, we investigate a range of boundary conditions that are relevant for viscous fluids~\cite{Batchelor_Fluid,Miles,KIDAMBI_2009,Gregory2025}.

Given the recent advances in using superfluid helium systems as gravity simulators~\cite{Svancara2024,Svancara2025}, we propose thin films of superfluid helium, a synthetic quantum system, for investigating mutual-information area laws. We present a framework for studying the mutual information of these systems, complemented by numerical predictions that display deviations from area-law scaling due to their inherent finite size.

In Section~\ref{section::theory}, we set out the theoretical formulation of quantum fields in the context of gravity simulators needed to estimate the covariance matrix from experimental measurements. We review measures of entropy and mutual information within Gaussian states obtainable from covariance matrices.
We then outline the applicability of our methods to the system of thin films of superfluid helium in Section~\ref{section:He thin films}, calling for the experimental investigation to be performed.
In Section~\ref{section::Results}, we show predictions for MI in experimentally plausible, $(2+1)$-dimensional systems in thermal states, including finite-size effects.
We finally discuss the implications of our methods, as well as future goals, in Section~\ref{section:Discussion}, with concluding remarks in Section~\ref{subsection::Conclusions}.

\section{Theory}\label{section::theory}
\vspace{-4mm}
A quantum field simulator is a physical system, whose effective degrees of freedom within a specific parameter regime may be identified with those of a quantum field~\cite{Unruh_1981,Liberati,Barroso_2023_shaker_th}. 
 Investigating these simulators can help us understand features of quantum fields through experimental realisations, testing the robustness of predictions even under non-idealised conditions. In this Section, we cover the necessary elements of quantum field theory before we investigate a concrete experimental platform in Section~\ref{section:He thin films}.
\subsection{\label{subsection::reconstruction theory} Quantum fields}
\vspace{-4mm}

We consider a $(d+1)$-dimensional quantum field $\hat{\phi}(t,\bm{x})$ in flat spacetime described by the Hamiltonian density
\begin{equation}\label{eq:general H}
    \hat{\mathcal{H}} ~=~ \frac{\hbar c}{2}\left[K|\nabla\hat{\phi}|^2 + \frac{1}{K}\hat{\eta}^2 + \frac{c^2}{\hbar^2}M^2K\hat{\phi}^2\right]\,,
\end{equation}where $c$ is the speed of wave propagation in the system, $M$ is the mass of the field $\hat{\phi}$, and $K$ is a system constant. The momentum conjugate to $\hat{\phi}$ is $\hat{\eta}$. The first term in~\eqref{eq:general H} is the kinetic energy of the field, the second is the momentum contribution, and the last term is the mass contribution. The field and the momentum are the system quadratures, obeying the equal-time canonical commutation relations (CCRs) $[\hat{\phi}(t, \bm{x}),\hat{\eta}(t, \bm{x}')]=\ii\delta^{(d)}(\bm{x}-\bm{x}')$. We note that we adopt the convention for field units such that there is no factor of $\hbar$ on the right-hand side of the CCRs. With this convention,~\eqref{eq:general H} recovers a Tomonaga-Luttinger Liquid Hamiltonian when $M=0$. The field and its momentum evolve under the massive Klein-Gordon equation,
\begin{equation}\label{eq:general EoMs}
    \left(\frac{1}{c^2}\partial_t^2 - \nabla^2 + \frac{c^2}{\hbar^2}M^2 \right)\hat{\phi}(t, \bm{x}) ~=~ 0\,.
\end{equation}

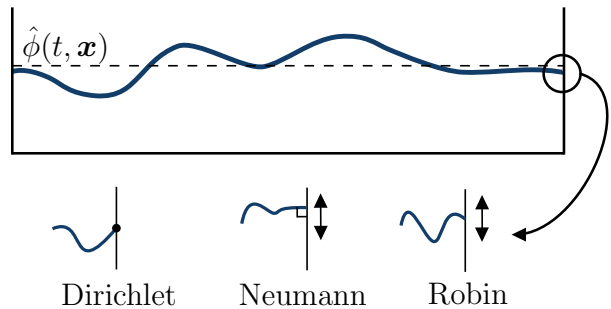
\begin{figure}
    \centering
    \hspace*{0\linewidth}
    \resizebox{1\linewidth}{!}{\tikzset{every picture/.style={line width=0.75pt}} 

\begin{tikzpicture}[x=0.75pt,y=0.75pt,yscale=-1,xscale=1]

\draw [color={rgb, 255:red, 19; green, 61; blue, 105 }  ,draw opacity=1 ][line width=2.25]    (110,89.67) .. controls (117.65,87.12) and (129.11,91.19) .. (135.17,94.5) .. controls (141.22,97.81) and (145.69,103.09) .. (161.17,104.5) .. controls (176.64,105.91) and (183.17,98) .. (188.17,93) .. controls (193.17,88) and (200.17,78.5) .. (212.17,74) .. controls (224.17,69.5) and (253.17,89) .. (265.67,86.5) .. controls (278.17,84) and (294.84,69.5) .. (311.67,68) .. controls (328.5,66.5) and (334,72.5) .. (342.5,76) .. controls (351,79.5) and (374.2,89.39) .. (388,90) .. controls (401.8,90.61) and (416.97,88.46) .. (431,88.5) .. controls (445.03,88.54) and (449.26,90.51) .. (449.83,89.83) ;

\draw[line width=1.2pt]   (110.33,50) -- (110.33,140.8) ;
\draw[line width=1.2pt]    (110.33,140) -- (450.33,140) ;
\draw[line width=1.2pt]    (450.33,50) -- (450.33,140.8) ;
\draw  [dash pattern={on 4.5pt off 4.5pt}]  (110.33,86) -- (450.33,86) ;

\draw[line width=1.2pt] (449.17,90.83) circle[radius=0.3cm];

\draw[line width=1.2pt] (459.9,90.83) edge [out=15, in=0] (420,190.17);

\draw [shift={(418,190.17)}, rotate = 0] [fill={rgb, 255:red, 0; green, 0; blue, 0 }  ][line width=0.08]  [draw opacity=0] (8.93,-4.29) -- (0,0) -- (8.93,4.29) -- cycle    ;

\draw    (174.5,160.67) -- (174.5,211.67) ;
\draw    (292,161.17) -- (292,212.17) ;
\draw    (389.5,162.17) -- (389.5,213.17) ;

\draw    (300.35,167.67) -- (300.48,190.67) ;
\draw [shift={(300.5,193.67)}, rotate = 269.67] [fill={rgb, 255:red, 0; green, 0; blue, 0 }  ][line width=0.08]  [draw opacity=0] (8.93,-4.29) -- (0,0) -- (8.93,4.29) -- cycle    ;
\draw [shift={(300.33,164.67)}, rotate = 89.67] [fill={rgb, 255:red, 0; green, 0; blue, 0 }  ][line width=0.08]  [draw opacity=0] (8.93,-4.29) -- (0,0) -- (8.93,4.29) -- cycle    ;

\draw    (399.85,168.67) -- (399.98,191.67) ;
\draw [shift={(399.83,194.67)}, rotate = 269.67] [fill={rgb, 255:red, 0; green, 0; blue, 0 }  ][line width=0.08]  [draw opacity=0] (8.93,-4.29) -- (0,0) -- (8.93,4.29) -- cycle    ;
\draw [shift={(399.83,165.67)}, rotate = 89.67] [fill={rgb, 255:red, 0; green, 0; blue, 0 }  ][line width=0.08]  [draw opacity=0] (8.93,-4.29) -- (0,0) -- (8.93,4.29) -- cycle    ;

\draw [color={rgb, 255:red, 19; green, 61; blue, 105}, draw opacity=1][line width=1.5]
(135.11,186.56) .. controls (156.78,177.22) and (144.5,221.67) .. (174.5,186.17);

\fill (174.5,186.17) circle[radius=2pt];

\draw [color={rgb, 255:red, 19; green, 61; blue, 105 }, draw opacity=1][line width=1.5] 
(251.73,183.4)
.. controls (256.13,160.6) and (267.29,177.09)
.. (271.73,177)
.. controls (276.18,176.91) and (270.67,173.08)
.. (291.5,173.5);

\draw [color={rgb, 255:red, 19; green, 61; blue, 105}, draw opacity=1][line width=1.5]
(349.89,184.56) .. controls (354.26,162.68) and (363.02,191.01) 
.. (369.44,194.22) .. controls (375.87,197.43) and (375.11,168.22) 
.. (389.5,180.67);

\draw    (285.92,173.08) -- (285.92,179.58) ;  
\draw    (285.92,179.08) -- (291.92,179.08) ;

\draw (115,60) node [anchor=north west][inner sep=0.75pt]   [align=left, font=\Large] {$\displaystyle \hat{\phi}(t,\bm{x}) \ $};
\draw (139,219) node [anchor=north west][inner sep=0.75pt]   [align=left, font=\Large] {Dirichlet};
\draw (248.5,219) node [anchor=north west][inner sep=0.75pt]   [align=left, font=\Large] {Neumann};
\draw (365,218) node [anchor=north west][inner sep=0.75pt]   [align=left, font=\Large] {Robin};

\end{tikzpicture}}
    \caption{\justifying Spatial profile of field $\hat{\phi}(t,\bm{x})$ satisfying Dirichlet, Neumann, or Robin boundary conditions. For Robin boundary conditions, the gradient and value of the field at the boundary are coupled by the parameter $\alpha$.}
    \label{fig:BCs}
\end{figure}

In the standard approach to field theory, fields are defined in spatially infinite regions. By contrast, experimental realisations typically simulate spatially confined effective fields. Accordingly, we work with fields confined within a finite volume $V$ and denote its boundaries by $\partial V$. The Hamiltonian~\eqref{eq:general H} and its associated equations of motion~\eqref{eq:general EoMs}, depend on the boundary conditions on the field, holding only for Dirichlet ($\hat{\phi}|_{\partial V}=0$) or Neumann ($\bm{n}\cdot\grad {\hat{\phi}}|_{\partial V}=0$) boundary conditions. Our analyses extend to a more general family of Robin boundary conditions (as encountered in viscous fluids~\cite{Batchelor_Fluid,Miles,KIDAMBI_2009,Gregory2025}), parametrised by a single real constant $\alpha$, $\bm{n}\cdot\grad \hat{\phi}|_{\partial V}=\alpha \hat{\phi}|_{\partial V}$ that recovers Dirichlet as $\alpha\to\infty$ and Neumann as $\alpha\to0$. We discuss Robin boundary conditions in greater detail in Appendix~\ref{app:Robin boundaries}. We depict the three boundary conditions in Figure~\ref{fig:BCs}.

A system is fully determined by both its equations of motions \textit{and} the boundary conditions; the boundary conditions alter the characteristics of the system as a whole, for example, by changing the density of states. For example, when considering the Neumann boundary condition, a so-called ``zero mode'' may appear, a zero-momentum mode (labelled zm) that contributes a constant offset in the field, which must be taken into account mathematically.

We consider now mode-sum solutions to \eqref{eq:general EoMs},
\begin{equation}\label{eq:general decomp}
    \begin{split}
        \hat{\phi}(t,\bm{x}) &~=~ \frac{1}{\sqrt{V}} \hat{\phi}_{\text{zm}}(t) + \sum_m \kappa_m^{\phi}g_m(\bm{x})\hat{\phi}_m(t)\,,\\
        \hat{\eta}(t,\bm{x}) &~=~ \frac{1}{\sqrt{V}} \hat{\eta}_{\text{zm}}(t) + \sum_m \kappa_m^{\eta}g_m(\bm{x})\hat{\eta}_m(t)\,,
    \end{split}
\end{equation}where $g_m$ are global, spatial mode functions, $\kappa_m^\phi$ and $\kappa_m^\eta$ are constant prefactors, and $\hat{\phi}_m$ and $\hat{\eta}_m$ are the mode operators, or momentum-space quadratures, obeying $[\hat{\phi}_m(t),\hat{\eta}_n(t)]=\ii \delta_{mn}$.

The spatial factors $g_m(\bm{x})$ satisfy the Helmholtz equation
\begin{equation}\label{eg:general Helmholtz}
    \nabla^2 g_m(\bm{x}) + k_m^2 g_m(\bm{x}) ~=~ 0\,,
\end{equation}
within $V$, where $k_m$ are the field momenta. The solutions to~\eqref{eg:general Helmholtz} depend on the conditions imposed on $g_m$ and its derivatives at the boundary $\partial V$ that describe the system's physical geometry. Given these boundary conditions, the set of solutions $g_m$ forms a complete and orthonormal set~\cite{Reed_Simon_2}.

To find the time evolution of the mode operators $\hat{\phi}_m$ and $\hat{\eta}_m$, we substitute the decomposition~\eqref{eq:general decomp} into the Hamiltonian density~\eqref{eq:general H} and integrate over the spatial volume, recovering the Hamiltonian operator $\hat{H}=\int_V\dd^{d}\bm{x}\,\hat{\mathcal{H}}$. We follow the conventions of \citet{Gluza2020}, in which the mode operators $\hat{\phi}_m(t)$ and $\hat{\eta}_m(t)$ are dimensionless, such that $\kappa_m^\phi = \sqrt{c/(K\omega_m)}$ and $\kappa_m^{\eta} = \sqrt{K\omega_m/c}$. For Dirichlet, Neumann, or Robin boundary conditions, the mode operators are described by the quantum harmonic oscillator Hamiltonian,
\begin{equation}\label{eq:general QHO}
    \hat{H} ~=~ \frac{\hbar\omega^{\eta}_{\text{zm}}}{2}\hat{\eta}_{\text{zm}}^2+\frac{\hbar \omega^{\phi}_{\text{zm}}}{2}\hat{\phi}^2_{\text{zm}}+\sum_m \frac{\hbar \omega_m}{2} \left[\hat{\phi}^2_m + \hat{\eta}^2_m \right]\,,
\end{equation} where the mode frequencies $\omega_m\coloneq\omega(k_m)$ follow the dispersion relation
\begin{equation}\label{eq:general disperion}
    \omega(k) ~=~ c \sqrt{k^2 + \frac{c^2}{\hbar^2}M^2}\,.
\end{equation} The zero modes, when present, are of frequencies $\omega^{\eta}_{\text{zm}}=c/K$ and $\omega^{\phi}_{\text{zm}}=c^3M^2K/\hbar^2$.

The mode operators evolve in time according to
\begin{equation}\label{eq:general mode t evoln}
    \begin{split}
        \hat{\phi}_m(t) &~=~ \hat{\phi}_m(0)\cos\lr{\omega_m t} + \hat{\eta}_m(0)\sin\lr{\omega_m t} \,, \\
        \hat{\eta}_m(t) &~=~ \hat{\eta}_m(0)\cos\lr{\omega_m t} - \hat{\phi}_m(0)\sin\lr{\omega_m t}  \,.
    \end{split}
\end{equation} We see clearly that the mode quadratures rotate into each other in phase space at frequency $\omega_m$. This follows from the fact that the time evolution of each quadrature depends on the initial conditions of both quadratures. This observation will form the basis of the methods that follow.

The dynamics of the zero modes are only harmonic for $M\neq0$, with frequency $\omega_{\text{zm}} = \frac{Mc^2}{\hbar}$. For $M=0$, we find
\begin{equation}
    \begin{split}
        \hat{\eta}_{\text{zm}}(t)~=~\hat{\eta}_{\text{zm}}(0)\,,\\
        \hat{\phi}_{\text{zm}}(t)~=~\omega_{\text{zm}}^{\eta}\hat{\eta}_{\text{zm}}(0)t + \hat{\phi}_{\text{zm}}(0)\,.
    \end{split}    
\end{equation} 

Hereafter, we will discuss reconstructing the covariance matrix of our two quadratures when the zero modes are not physical observables. A reconstruction scheme has been reported in~\cite{Tajik_MI_2023, TajikPhD} for ultra-cold atoms systems with physically observable zero modes.

We finally note that one may consider implementing a quench --- a rapid change of the Hamiltonian --- to study thermal states generated by various initial Hamiltonians, because the reconstruction procedure recovers the statistics of the initial state. A quench also amplifies the signal of the system quadratures at the moment of quenching, improving the signal-to-noise ratio.  Moreover, as a quench specifies an initial time, it has been shown that one can experimentally probe the time evolution of a system by a series of repeated reconstruction procedures~\cite{Aimet_2024_Landauer}.

\subsection{Covariance reconstruction}\label{subsection::Covariance reconstruction}
\vspace{-4mm}
We now apply the formalism outlined in Section~\ref{subsection::reconstruction theory} to show how Gaussian states may be fully reconstructed from experimental measurements. As such, coordinates $\bm{x}$ are now a set of discretised coordinates.

When the state of the system is Gaussian, it is fully specified by the first and second statistical cumulants of the system quadratures $\hat{\phi}(t, \bm{x})$ and $\hat{\eta}(t, \bm{x})$, which obey the canonical commutation relations (CCRs) $[\hat{\phi}(t, \bm{x}_i),\hat{\eta}(t, \bm{x}_j)]=\ii \delta_{ij}/V$. For reasons that will be made clear in Section~\ref{subsection::Gaussian Info}, we focus on the second cumulant, the covariance matrix
\begin{equation}\label{eq:covariance blocks}
    \Gamma ~=~ \begin{pmatrix}
        Q & R \\
        R^{\text{T}} & P
    \end{pmatrix}\,,
\end{equation}where $Q$, $R$, and $P$ are matrices of the quadrature elements, $Q_{ij}=\langle\hat{\phi}(t, \bm{x}_i)\hat{\phi}(t, \bm{x}_j)\rangle$, $R_{ij} = \frac{1}{2} \langle\{ \hat{\phi}(t, \bm{x}_i), \hat{\eta}(t, \bm{x}_j) \}\rangle$, and $P_{ij} = \langle\hat{\eta}(t, \bm{x}_i) \hat{\eta}(t, \bm{x}_j)\rangle$. The anti-commutator is denoted $\{\cdot,\cdot\}$. To extract the covariance matrix experimentally, one requires simultaneous access to both system quadratures --- a formidable task, especially in continuous systems. Experimentally, one can typically access one of the two-point functions $\langle\hat{\phi}(t,\bm{x}_1)\hat{\phi}(t,\bm{x}_2)\rangle$ or $\langle\hat{\eta}(t,\bm{x}_1)\hat{\eta}(t,\bm{x}_2)\rangle$. If a high-level modelling of the system is in place, one may predict the time evolution of the two-point function (see Appendix~\ref{app:reconstruction}) in terms of $\tilde{Q}_{mn}(0)$, $\tilde{P}_{mn}(0)$, and $\tilde{R}_{mn}(0)$, where
\begin{equation}\label{eq:QPR momentum}
\begin{split}
    \tilde{Q}_{mn}(t)~&=~\langle\hat{\phi}_m(t)\hat{\phi}_n(t)\rangle\,,\\
    \tilde{P}_{mn}(t)~&=~\langle\hat{\eta}_m(t)\hat{\eta}_n(t)\rangle\,,\\
    \tilde{R}_{mn}(t)~&=~\frac12\langle\{\hat{\phi}_m(t)\hat{\eta}_n(t)\}\rangle\,,
\end{split}
\end{equation}are blocks of the momentum-space covariance matrix $\tilde{\Gamma}$. These momentum-space covariance elements can be fitted from experimental data using matrix linear regression of~\eqref{eq:general reconstruction phiphi} or~\eqref{eq:general reconstruction etaeta}. Whilst this process reconstructs the covariance matrix blocks for the momentum-space quadratures~\eqref{eq:QPR momentum}, one may also study the system's spatial properties, by transforming to real space through a discrete transformation compatible with $g_m(\bm{x})$.

We note that, in practice, the finite resolution of any measurement or numerical simulation results in a finite set of modes being reconstructed. As a consequence, to preserve the CCRs, both quadratures are rescaled as $\phi\rightarrow \phi/\epsilon$, where $\epsilon$ is the unit cell lattice volume. The number of successfully reconstructed modes determines the effective resolution of the discretised spatial grid of the final reconstruction.

The above procedure formalises the extraction of covariance from experimental realisations of simulator systems. Having obtained the covariance matrix, the Gaussian state is fully characterised, enabling studies of correlations and entanglement. In the next Section, we focus on a particularly important quantifier of correlations, intimately linked to fundamental physics --- mutual information.

We bring this Section to a close by remarking that, whilst we have focussed on Gaussian states, one may still appeal to this procedure for non-Gaussian states and reconstruct the covariance matrix.  Although the covariance matrix is insufficient for fully describing the state of the system, it may be used to calculate a lower bound of the mutual information~\cite{Cover_thomas_2006,Wilde_2013}.

\subsection{Information theory for Gaussian states}\label{subsection::Gaussian Info}
\vspace{-4mm}

From a general state of a quantum system described by a density matrix $\rho$, one may calculate the von Neumann entropy
\begin{equation}\label{eq::vN_S_defn}
    S\lr{\rho}= - \text{Tr}\left[\rho\log\rho\right] \, .
\end{equation}For a composite system with total Hilbert space $\mathscr{H} = \mathscr{H}_A \otimes \mathscr{H}_B$, comprising two subsystems $A$ and $B$, the state of subsystem $A$ is given by the partial trace of $\rho$ over system $B$, $\rho_{A}=\Tr_{B}[\rho]$, and vice-versa for the state of subsystem $B$. Given these states, one may calculate the \textit{quantum mutual information} (MI),
\begin{equation}\label{eq:MI_defn}
    I\lr{A : B} ~=~ S\lr{\rho_A} + S\lr{\rho_B} - S\lr{\rho}\,.
\end{equation}

We remark that both MI $I(A:B)$ and von Neumann entropy $S(\rho)$ quantify entanglement in pure states (for which $\Tr[\rho^2]=1$). For mixed states ($\Tr[\rho^2]<1$), however, they fail to distinguish entanglement from classical correlations. Nevertheless, MI is a robust measure of correlations --- whether classical or quantum --- between subsystems $A$ and $B$~\cite{Wilde_2013, Zurek2001, Adesso2007, Adesso_2010}.

In general, one must reconstruct the state $\rho$ from experimental measurements in order to calculate either the von Neumann entropy or MI. Gaussian states, however, are fully characterised by the first and second cumulants; furthermore, only the second cumulant is used in calculating any information-theoretic quantity. For this class of states, we need only consider the state covariance matrix $\Gamma$ described in Section~\ref{subsection::Covariance reconstruction} to compute the von Neumann entropy,
\begin{equation}\label{eq::Gaussian Entropy}
    S(\Gamma)\!=\!\sum_{n=1}^N(\gamma_n+\tfrac12)\log(\gamma_n+\tfrac12) - (\gamma_n-\tfrac12)\log(\gamma_n-\tfrac12)\,,
\end{equation}where $\gamma_n$ are the \textit{symplectic eigenvalues} of $\Gamma$, i.e., the eigenvalues of $\ii\Omega\Gamma$, where $\Omega$ is the symplectic block matrix~\cite{Eisert2010, Demarie2018, Adesso_2014},
\begin{equation}
    \Omega~=~\begin{pmatrix}
        0 & 1_{N\times N}\\
        -1_{N\times N}& 0
    \end{pmatrix}\,.
\end{equation}

\begin{figure}
    \centering
    \hspace*{-0.11\linewidth}
    \resizebox{1.\linewidth}{!}{\input{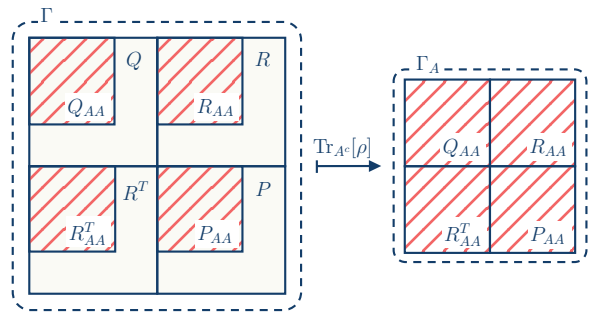}}
    \caption{\justifying Structure of the covariance matrix $\Gamma$~\eqref{eq:covariance blocks} and visualisation of the partial trace, forming the covariance matrix of subsystem $A$.}
    \label{fig:Covariance_Matrix}
\end{figure}

To calculate information between subsystem $A$ and its complement $A^c$, we must understand the operation of partial tracing on Gaussian states. First, each of the matrices $Q$, $R$, and $P$ in~\eqref{eq:covariance blocks} are block matrices of the form
\begin{equation}
    Q~=~\begin{pmatrix}
        Q_{AA}&Q_{AA^{c}}\\
        Q^{\text{T}}_{AA^c}& Q_{A^cA^c}
    \end{pmatrix}\,,
\end{equation}where $Q_{AA}$ and $Q_{A^cA^c}$ are matrices of quadratures in $A$ and $A^c$ respectively, while $Q_{AA^c}$ contains all of the terms common to $A$ and $A^c$. These blocks partition $\Gamma$ into covariance matrices $\Gamma_A$ and $\Gamma_{A^c}$; the behaviour of system $A$ is fully described by
\begin{equation}
    \Gamma_A~=~\begin{pmatrix}
        Q_{AA}&R_{AA}\\
        R_{AA}^{\text{T}}&P_{AA}
    \end{pmatrix}\,.
\end{equation}This partitioning of covariance matrices $\Gamma$ and $\Gamma_A$ (depicted in Fig.~\ref{fig:Covariance_Matrix}) is equivalent to taking the partial trace $\Tr_{A^c}[\rho]$. The entropy of each subsystem may then be computed using~\eqref{eq::Gaussian Entropy}, and hence MI may be computed using~\eqref{eq:MI_defn}.

As an example, consider a freely evolving system. Here, mode-mode interactions are absent, and MI is trivially zero. By contrast, we find a non-zero MI after transforming the covariance matrix to real space $\Gamma$. Whilst initially surprising, this may be understood by remarking that the transformations to real space involve a sum over all momentum modes and a redefinition of subsystems $A$ and $A^c$.

We highlight that the outlined procedures enable the experimental studies of MI; once the momentum-space covariance matrix $\tilde{\Gamma}$ has been obtained (see~\ref{subsection::Covariance reconstruction}), one may then compute MI between momentum modes or spatially extended regions. Furthermore, because the covariance elements describe the initial state, if a quench is performed at $t=t_0$, the procedures in this Section enable the study of final Gaussian states created by a range of configurations as well as the time evolution of information. For more details, we defer the reader to~\cite{Aimet_2024_Landauer}. In the next Section, we consider a concrete experimental simulator for investigating information in quantum field theory.

\section{\label{section:He thin films} Proposed simulator}
\begin{figure}
    \centering
    \includegraphics[width=0.8\linewidth]{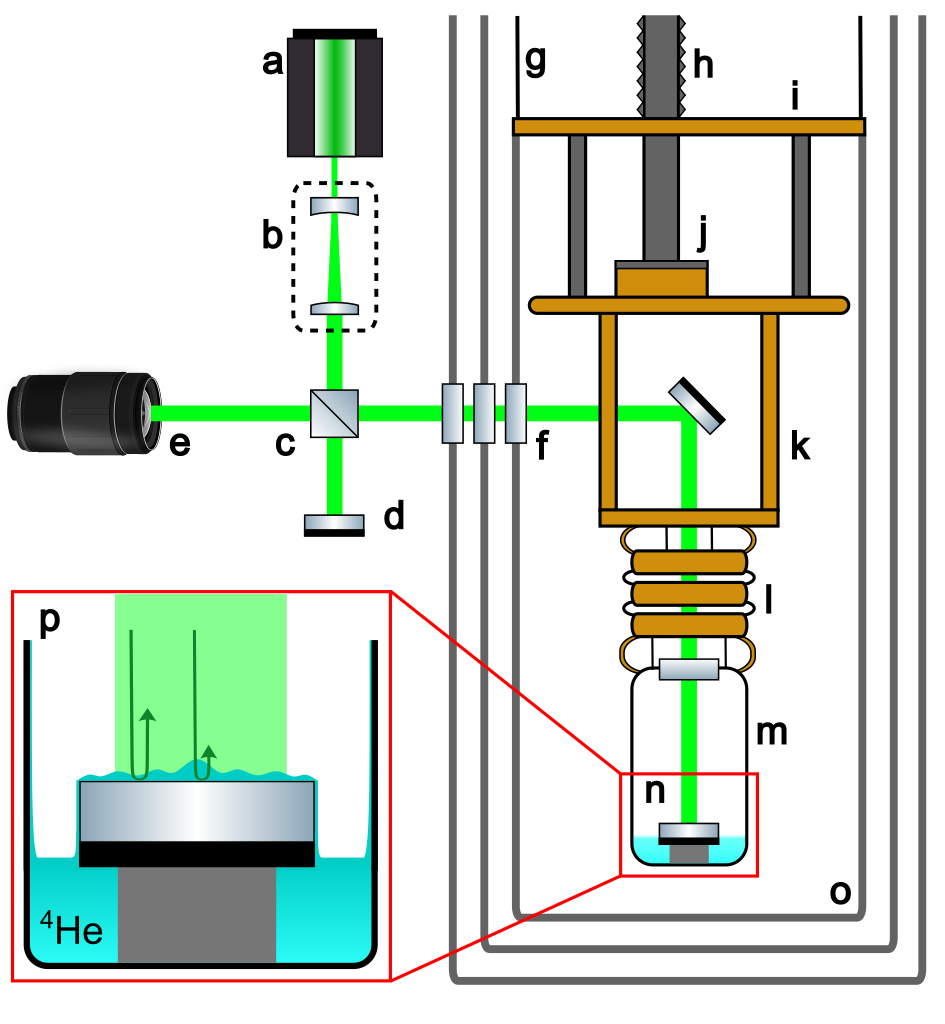}
    \caption{\justifying Proposed experimental setup for measuring MI in thin films of superfluid helium using digital holography in a $^3$He refrigerator with internal vibrational shielding. $\textbf{(a)}$ Laser, $\textbf{(b)}$ beam expander, $\textbf{(c)}$ beam splitter, $\textbf{(d)}$ reference-beam mirror, $\textbf{(e)}$ digital camera, $\textbf{(f)}$ cryostat optical-access windows, $\textbf{(g)}$ suspension wires, $\textbf{(h)}$ flexible bellows, $\textbf{(i)}$ intermediate cold plate, $\textbf{(j)}$ 0.3K $^3$He pot $\textbf{(k)}$ thermal link to experimental cell, $\textbf{(l)}$ low-pass vibration filter multi-pendulum stage, $\textbf{(m)}$ experimental cell, $\textbf{(n)}$ target-beam mirror, $\textbf{(o)}$ radiation shields, $\textbf{(p)}$ inset showing phase fluctuations imprinted on the laser by thin-film height fluctuations $\delta h$.}
    \label{fig:exp_setup}
\end{figure}

\vspace{-4mm}
We examine now an experimental platform not previously considered for the purpose of investigating quantum information in the context of field-theory simulations: thin films of superfluid helium-$4$~\cite{Atkins1959,Baker2016, Bunney2023_published}. When placed within a cryogenic cell, superfluid helium (helium-II) will coat the surfaces of its enclosure, forming an interface between the liquid and vapour phases. In a confined geometry, the normal component of helium-II is viscously clamped to the substrate, whilst the superfluid component can move freely~\cite{Atkins1959}. Consequently, extremely thin films of superfluid form with thicknesses from a few atomic layers up to hundreds of nanometres~\cite{Baker2016}. The van der Waals interactions between helium atoms and the substrate material dominate the dynamics of the thin-film interface. The equilibrium thickness $h_0$ of such thin films is determined by the physical and geometric properties of the substrate and the conditions in the cryogenic cell~\cite{sabisky1973onset}. A schematic diagram depicting a superfluid helium cryostat with optical access for non-destructive, space and time-resolved measurements of thin films of superfluid helium is shown in Fig.~\ref{fig:exp_setup}.

The inviscid propagation of surface waves on superfluid thin films may be described well by the dispersion relation~\cite{reeves_2025, rochePRL1995, sfendla2021extreme}
\begin{equation}\label{eq:thin film dispersion}
    \omega^2(k) ~=~ \geff\lr{1+\ell_c^2k^2}k\tanh(kh_0)\,,
\end{equation}where $k$ is the wavenumber, $\geff = 3\alpha_{\text{vdW}} h_0^{-4}$ is the van der Waals-dominated effective gravity, and $\alpha_{\text{vdW}}$ is the van der Waals coefficient between helium atoms and the substrate~\cite{rochePRL1995}. The capillary length $\ell_c^2=\sigma/(\rho \geff)$ is the characteristic length scale at which the surface tension $\sigma$ between the film and helium vapour dominates the dispersion relation. For sufficiently long wavelengths ($kh_0\ll 1$ and $k\ll \ell_c^{-1}$), these surface waves satisfy a linear dispersion relation, $\omega(k) = c_3k$ with $c_3=\sqrt{g_{\text{eff}}h_0}$, and are commonly referred to as third-sound waves~\cite{Atkins1959}.

The confinement size determines the available wavenumbers $k$, whilst the fluid depth controls the speed of propagation $c_3$, and a combination of both determines the mode frequencies $\omega$ through~\eqref{eq:thin film dispersion}. Currently available, experimental systems range in size from micro-cavity resonators~\cite{reeves_2025, sfendla2021extreme, Sachkou_2019} to centimetre-sized cells~\cite{ketola1992anomalous, herrmann1998, herrmann1999}, both allowing precise control of the film thickness $h_0$. This allows the tuning of the frequencies in the system across $1$Hz--$1$MHz for the lowest-lying excitations, the smallest $\omega$, at propagation speeds $1$--$10^{-3}$ms${}^{-1}$. For instance, in a centimetre-scale experimental cell, conditions may be readily achieved to create a thin film with, say, $T=0.3$K and $h_0=80$nm. When developed on a sapphire substrate with $\alpha_{\text{vdW}} \approx 2.6 \times 10^{-24}\mathrm{m}^{5}\mathrm{s}^{-2}$, we have $c_3\approx 0.12$ms$^{-1}$, upon which an effective field theory may be defined.

When considering third sound, the dynamics of the surface degrees of freedom can be described by a linearised theory for perturbations of the height $\delta h$ and the irrotational velocity potential $\phi$. In this regime, these follow equations of motion governed by the Hamiltonian density
\begin{equation}\label{eq:He KG H}
    \mathcal{H} ~=~ \frac{1}{2}\rho \left[h_0 |\nabla \phi|^2 + \geff\delta h^2\right] \, ,
\end{equation}
where $\rho$ is the superfluid density. Excitations on the helium-vapour interface, known as third-sound phonons, may be described by the quantised quadratures $(\hat{\phi}, \delta \hat{h})$ that follow the CCR $[\hat{\phi}(t, \bm{x}), \rho \delta \hat{h}(t, \bm{x}')] = -\ii\hbar\delta\lr{\bm{x} - \bm{x}'}$~\citep{Landau1941,Bunney2023_published,Cameron_thesis}. 

Using a Madelung representation, we consider instead phase and number density fluctuations, $\hat{\varphi} = \hbar m_4^{-1}\hat{\phi}$ and $\hat{\pi} = -\rho m_4^{-1} \delta\hat{h}$, respectively, where $m_4$ is the mass of a helium-$4$ atom. These quadratures satisfy the CCR $[\hat{\varphi}(t, \bm{x}), \hat{\pi}(t, \bm{x}')]=\ii \delta(\bm{x}-\bm{x}')$. This maps our theory onto the Tomonaga-Luttinger Liquid (TLL) Hamiltonian~\eqref{eq:general H} with $M=0$,
\begin{equation}\label{eq: He TLL H}
    \hat{\mathcal{H}} ~=~ \frac{\hbar c_3}{2}\left[K |\bm{\nabla} \hat{\varphi}|^2 + \frac{1}{K}\hat{\pi}^2\right],
\end{equation}with Luttinger parameter $K=\hbar \rho c_3/(\geff m_4^2)$, approximately $2.21\times 10^{14} \mathrm{m}$, based on the experimental parameters above.

We note that the zero modes do \textit{not} contribute to the observable dynamics. Our degrees of freedom are linearised around a background, and the zero modes create spatially homogeneous offsets in the velocity potential and the film height; hence, the relevant degrees of freedom are unaffected --- zero modes are not physically observable.

Following the correspondence between the effective Hamiltonian density describing third-sound excitations and the TLL Hamiltonian~\eqref{eq:general H} with $M=0$, we identify the phase $\hat{\varphi}$ as an effective $(2+1)$-dimensional massless quantum field, with conjugate momentum $\hat{\pi}$, to which we may directly apply the formalism of in Section~\ref{section::theory}. Consequently, if one can perform space and time-resolved measurements of one of the quadratures, one can reconstruct the covariance matrix to study MI of third-sound waves.

Imaging methods, such as digital holography (see Fig.~\ref{fig:exp_setup}) and Fourier checkerboard demodulation~\cite{Barroso2023, Wildeman_2018, Svancara2024}, can be employed to reconstruct the three-dimensional profile of the free surface of superfluid helium over time. Thus, the height perturbations $\delta h \lr{t, \bm{x}}$ are reconstructed on a spatial grid of points and discretely sampled over time, limited by the specifications of the setup, e.g., the camera sample rate or the imaging system resolution.
The extension of these methods to the thin-film regime would recover the two-point function of conjugate momentum, and enable the measurement of MI, as specified in Section \ref{section::theory}.


It may also be possible to engineer scenarios where third-sound excitations in thin films of superfluid helium could become effectively massive, for example, by employing narrow transverse trapping. This results in an effective Kaluza-Klein-like tower of masses~\cite{Leizerovitch:2017cge}. With sufficient trapping, the massive branches separate in energy. Due to this trapping, however, the resultant simulator is $(1+1)$-dimensional. Other methods may also be applicable, such as imposing strong, fast electromagnetic field oscillations at the interface.

\begin{figure*}[ht]
    \centering
    \hspace*{-0.065\linewidth}
    \resizebox{1.15\linewidth}{!}{\input{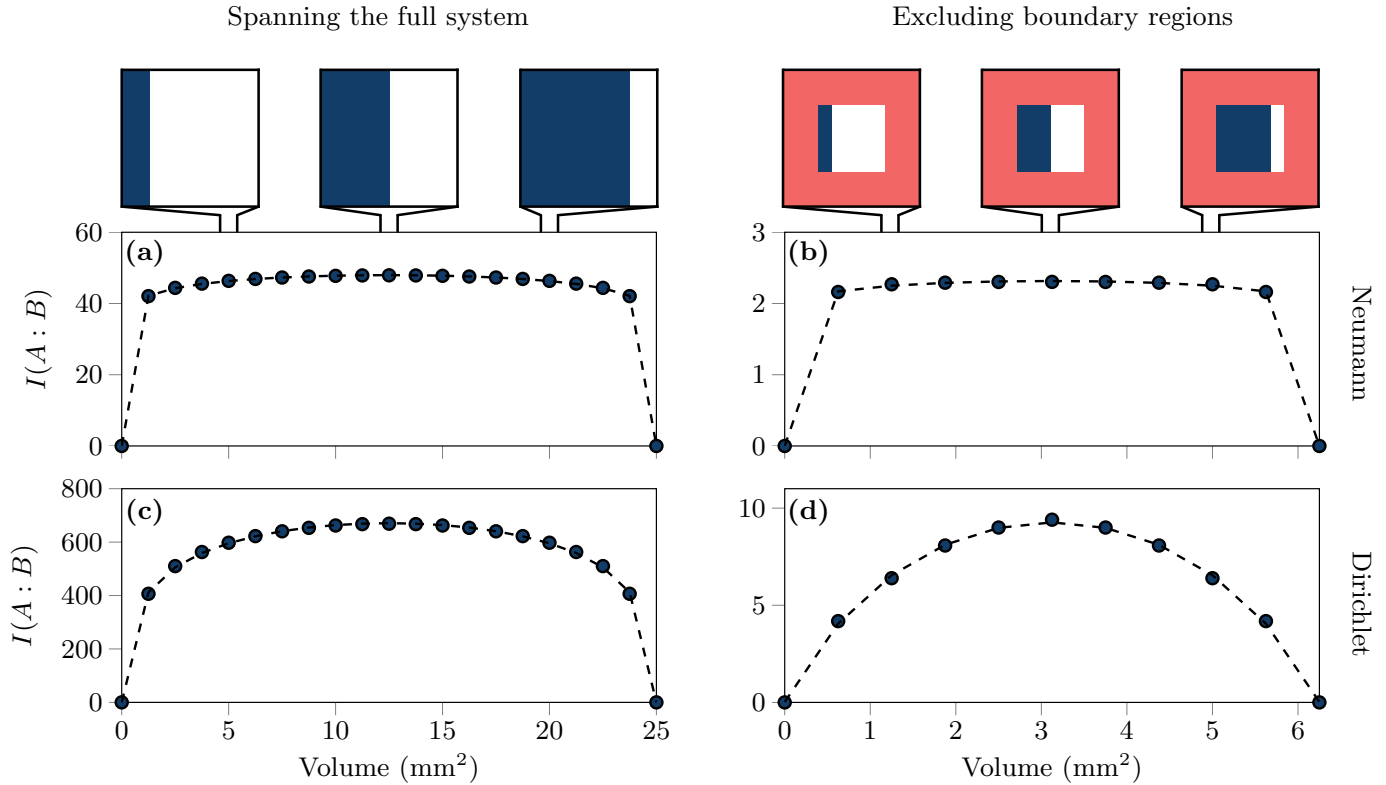}}
    \caption{
    \justifying Mutual information prediction in thin-film superfluid helium-$4$ with varying subsystem volume with constant boundary area. Simulation parameters: $h_0=80$nm, $T=0.3$K, $c_3=0.12$ms$^{-1}$, $K=2.21\time 10^{14}$m, $L_x=L_y=5$mm, $N_x=N_y=20$, and $\Delta x=\Delta y=0.25$mm. Insets depict subsystems $A$ (navy) and $B$ (white). \textbf{(a)} Mutual information with varying volume of subsystem $A$ with Neumann boundary conditions. Dashed line fitted prediction~\eqref{eq:Calabrese fit} from~\cite{Calabrese2004}. \textbf{(b)} Mutual information with varying volume of subsystem $A$ far from cell boundary with Neumann boundary conditions. Dashed line fitted prediction~\eqref{eq:Calabrese fit} from~\cite{Calabrese2004}.
    \textbf{(c)} Mutual information with varying volume of subsystem $A$ with Dirichlet boundary conditions. Dashed line fitted prediction~\eqref{eq:Calabrese fit} from~\cite{Calabrese2004}.
    \textbf{(d)} Mutual information with varying volume of subsystem $A$ far from cell boundary with Dirichlet boundary conditions. Dashed line fitted prediction~\eqref{eq:Calabrese fit} from~\cite{Calabrese2004}.}
    \label{fig:MI_vs_V}
\end{figure*}

\section{\label{section::Results} Results}
\vspace{-4mm}

We present now an illustrative example of information distributions in a $(2+1)$-dimensional scalar field that is subject to spatial confinement. A particular experimental platform for investigating this is thin-film superfluid helium, as discussed in Section~\ref{section:He thin films}. Until now, we have kept the state of the quantum field general. We specialise now to a thermal state, modelling the low but non-zero temperature of a sample of superfluid helium. The momentum-space covariance matrix elements are
\begin{equation}\label{eq::mode covar thermal}
    \begin{split}
        \tilde{Q}_{mn} ~&=~ \left(n_{T}(\omega_m) + \frac{1}{2}\right)\delta_{mn}\,, \\
        \tilde{P}_{mn} ~&=~ \left(n_{T}(\omega_m) + \frac{1}{2}\right)\delta_{mn}\,, \\
        \tilde{R}_{mn} ~&=~ 0 \,,
    \end{split}    
\end{equation}where $n_T(\omega_m)=(\exp[\hbar\omega_m/(k_{\text{B}T})]-1)^{-1}$ is the Bose-Einstein distribution at temperature $T$.

To study area laws within this setting, we must first construct the spatial covariance matrix from~\eqref{eq::mode covar thermal}. To do this, we must know the system's geometry, and hence boundary conditions. We consider a field confined within a rectangular geometry with Dirichlet (Neumann) boundary conditions, in which the spatial factors $g_m$ are sine (cosine) functions. By performing a discrete sine (cosine) transformation~\cite{SciPy_2020, Makhoul_1980} of the momentum-space covariance matrix elements~\eqref{eq::mode covar thermal}, we recover the spatial covariance matrix elements $(Q,R,P)$. We then use these elements to construct the covariance matrix $\Gamma$, and study the MI between various spatial regions.

We note that, when dealing with Neumann boundary conditions, we must consider the zero-modes with care; in simulators of massless field theories ($M=0$ in~\eqref{eq:general H}), the zero modes cause a divergence in MI~\cite{Aimet_2024_Landauer}. As discussed in Section~\ref{section:He thin films}, however, the zero modes in our proposed simulator are not physical degrees of freedom. Hence, there is no zero-mode contribution in~\eqref{eq::mode covar thermal}.

\begin{figure*}[ht]
    \centering
    \hspace*{-0.04\linewidth}
    \resizebox{1.11\linewidth}{!}{\input{figures/tikz/Fig_3_tikz.tex}}
    \caption{\justifying Mutual information prediction in thin-film superfluid helium-$4$ with varying boundary area at constant subsystem volume. Simulation parameters: $h_0=80$nm, $T=0.3$K, $c_3=0.12$ms$^{-1}$, $K=2.21\time 10^{14}$m, $L_x=L_y=5$mm, $N_x=N_y=20$, and $\Delta x=\Delta y=0.25$mm. Insets depict subsystems $A$ (navy) and $B$ (white).
    \textbf{(a)} Mutual information with varying boundary area between subsystems $A$ and $B=A^c$ with Neumann boundary conditions.
    \textbf{(b)} Mutual information with varying boundary area between subsystems $A$ and $B=A^c$ far from the cell boundary with Neumann boundary conditions.
    \textbf{(c)} Mutual information with varying boundary area between subsystems $A$ and $B=A^c$ with Dirichlet boundary conditions.
    \textbf{(d)} Mutual information with varying boundary area between subsystems $A$ and $B=A^c$ far from the cell boundary with Dirichlet boundary conditions.
    }
    \label{fig:MI_vs_A}
\end{figure*}


In one spatial dimension, it is sufficient to verify the area law of MI by confirming that the MI does not respond to changes in volume~\cite{Tajik_MI_2023}. By contrast, it is a necessary --- but not sufficient --- condition in two spatial dimensions that the MI be volume-insensitive~\cite{Wolf2008, Casini_2009, Eisert2010, casini2023_Et_in_QFT}. The classification of area laws further requires MI to scale, at most, proportionally to the area of the boundary~\cite{Wolf2008}.

To investigate MI scaling laws, we study the distribution of two-dimensional MI, using the theoretically constructed spatial covariance matrix~\eqref{eq::mode covar thermal}. We compute MI in a rectangular geometry for various subsystems $A$ and $B$, considering Dirichlet and Neumann boundary conditions separately. In Fig.~\ref{fig:MI_vs_V} and Fig.~\ref{fig:MI_vs_A}, we consider a rectangular cell with side lengths $L_x=L_y=5$mm, spatial resolution $N_x=N_y=20$, and pixel dimensions $\Delta x=\Delta y=0.25$mm. We consider thin films of superfluid helium with $h_0=80$nm, $T=0.3$K, and $c\approx 0.12$ms$^{-1}$. For consistency, we ensure a constant pixel separation in defining subsystems $A$ and $B$.

In Fig.~\ref{fig:MI_vs_V}, we study how MI varies with changes in subsystem volume with a constant boundary area. Subsystems $A$ and $B$, depicted in the insets, are defined by a vertical division, whose horizontal displacement determines the subsystem volume. In Fig.~\ref{fig:MI_vs_V}~(a) and (c), subsystems $A$ and $B=A^c$ span the entire cell to investigate boundary effects. In~Fig.~\ref{fig:MI_vs_V}~(b) and (d), subsystems $A$ and $B$ are defined away from the exterior boundary.

We consider Neumann boundary conditions in Fig.~\ref{fig:MI_vs_V}~(a) and (b). As expected, we observe volume-insensitivity, a necessary signature of an area law. Far from the cell boundary, any deviations from a flat curve are further diminished (Fig.~\ref{fig:MI_vs_V}~(b)). 

In Fig.~\ref{fig:MI_vs_V}~(c) and (d), we consider Dirichlet boundary conditions. Whilst the MI follows a qualitatively similar behaviour, it exhibits a more prominent sensitivity to volume. Unlike the case of Neumann boundary conditions, the volume contribution to the MI are \textit{not} suppressed by separating subsystems $A$ and $B$ from the cell boundary (Fig.~\ref{fig:MI_vs_V}~(d)). We may understand this by considering the distribution of information locally in the system; concretely, we define subsystem $A$ to be a single pixel, with $B=A^c$. By analysing each pixel in the system in turn, we map out the local information (see Appendix~\ref{app:MI maps}). In the case of Dirichlet boundary conditions, local information varies most near the cell boundaries, whilst remaining almost constant near the centre. Therefore, we expect a qualitatively different MI behaviour between including and excluding the exterior boundary. For Neumann boundary conditions, however, the variation of local information is uniform throughout the cell, leading to the observed cell boundary-insensitivity in (a) and (b).

Comparing the magnitude of MI in the case of Neumann and Dirichlet boundary conditions, we see that the MI between subregions in a cell with Dirichlet boundary conditions is always larger than that with Neumann boundary conditions. This is a direct consequence of removing the zeros modes, which would otherwise contribute significantly to the overall information content.

We note that, though we made these predictions for highly mixed states, the MI curves in Fig.~\ref{fig:MI_vs_V} can be well fitted using predictions of finite-size area laws for the entropy of \textit{pure} states~\cite{Calabrese2004}. Whilst originally derived using replica methods~\cite{casini2023_Et_in_QFT} for one-dimensional conformal field theories with open boundary conditions, we may lift their result to two-dimensional field theories,
\begin{equation}\label{eq:Calabrese fit}
    S ~\approx~ \kappa_1 \log\Bigl[ \frac{N_xN_y}{\pi}\sin\lr{\frac{\pi V}{\Delta x \Delta y}} + \kappa_2 \Bigr] + \kappa_3\,,
\end{equation} where $\kappa_i$ are fitting parameters, and $V$ is the $2$D cell volume. In~\cite{Calabrese2004}, $\kappa_1$ is related to the conformal charge in the conformal field theory. Despite the considerable differences between the two systems, we note that these predictions have been conjectured to apply in more general settings, and have been employed in~\cite{Deller2024a, Deller2024b} to measures of MI. Should a direct link between these predictions and MI be proven, our results will pave the way for experimental measurements of conformal charges in analogue systems.

In Fig.~\ref{fig:MI_vs_A}, we investigate how MI varies with changes in boundary area, whilst maintaining a constant subsystem volume. We depict this procedure in the insets in Fig.~\ref{fig:MI_vs_A}. Again, we consider both Neumann and Dirichlet boundary conditions, and the effect of including or excluding the cell boundary. As a technical aside, we ensure to keep the number of corners in each subsystem constant when varying the boundary area to avoid additional UV divergences in the subsystem entropy~\cite{casini2023_Et_in_QFT}.

For both Neumann and Dirichlet boundary conditions, whether including or excluding the cell boundary, we find at most a linear scaling of MI with the boundary area. In combination with volume-insensitivity, this is sufficient for verifying area-law scaling of mutual information.

The magnitude of MI is again significantly larger in the case of Dirichlet boundary conditions, reaching a maximum value of approximately $1.1\times10^3$ nats (natural units of information). Finally, we note in Fig.~\ref{fig:MI_vs_A}~(c) a breaking in the monotonicity of MI scaling; however, this is remedied by excluding the cell boundary.

\section{\label{section:Discussion} Discussion and Outlook}
\vspace{-4mm}
The predictions presented in Section~\ref{section::Results} indicate that the observation of the area law of mutual information is in principle possible. To achieve this, we identify four key tasks required to fully characterise mutual information in the system of interest: measurement of height fluctuations with spatio-temporal resolution; Gaussianity of the state of the effective fields; knowledge of boundary conditions; and linear evolution of emerging degrees of freedom within required timescales.

As discussed in Section~\ref{section:He thin films}, spatio-temporally resolved measurements of the effective conjugate momentum, height perturbations, have been achieved, resolving sub-micrometric changes on the fluid-vapour interface of superfluid helium samples with depths ranging from hundreds of microns to a few centimetres~\cite{Barroso2025}. On the other hand, we expect interfacial changes below a few nanometres in thin films of superfluid helium~\cite{reeves_2025}, which should be achievable with optical imaging methods, such as digital holography~\cite{huang2024quantitative, kumar2023emerging}, appropriately adapted to cryogenic setups. 

The assumption that the state of the emergent degrees of freedom is a Gaussian thermal state requires the occupation of interfacial modes to be well described by a thermal spectrum. In an experiment, environmental, mechanical noise may dominate within specific frequency ranges, therefore placing strict requirements on the noise-isolation threshold.

The experiment can be shielded from environmental noise using commercially available cross-braced stiffened frames, commonly used as supporting structures for dilution refrigerators. Additional damping mechanisms can then be added to shield the noise generated by pulse-tube cryo-coolers that drive the primary cooling within the fridge. This can be achieved by mechanically decoupling the cold stages below 1K~\cite{caparrelli2006vibration} and/or by creating a low-pass vibration filter with high-thermal-conductance coupled-pendula~\cite{de2019vibration}, allowing shielding within the frequency range of interest, in which thermal effects may be resolvable. 

The methods outlined in Section~\ref{subsection::Covariance reconstruction} only require the linear evolution of the effective fields, and hence extend beyond thermal states. Therefore, if the remaining noise sources dominate above thermal effects, our methods still directly apply, allowing studies of information in non-thermal states, provided the noise profile is sufficiently close to Gaussian. Contrary to this, if the experimental realisations are characterised by non-Gaussian statistics, reconstruction of the full state of a continuous quantum field simulator is required. This is beyond current experimental reach, with state-of-the-art tomographic methods reconstructing on the order of tens of qubits~\cite{Lanyon:2017uue,Torlai:2018}. Although the state itself does not affect the reconstruction methods of Section~\ref{subsection::Covariance reconstruction}, this procedure provides only the second cumulant, with even higher-order cumulants being insufficient to quantify entropy measures for non-Gaussian states. Nonetheless, these methods place lower bounds on the mutual information in systems with non-Gaussian statistics~\cite{Cover_thomas_2006}.

We recall that, in our work, we need only measure a single quadrature. As such, only the boundary conditions for the measured quadrature --- and not both --- must be specified. The spatial resolution of available measurement schemes~\cite{Barroso2023} enables the boundary conditions to be immediately inferred~\cite{Barroso2025, Baker2016}. Additionally, the system boundary conditions may be experimentally controllable by: altering the cryogenic substrate of the boundary region; changing the geometry of the confining material; or deploying a barrier whose surface thin-film helium will not wet. Boundaries made of caesium have precisely the last property, with the additional advantage of providing a reflective boundary to third-sound waves~\cite{ketola1992anomalous}.

After measuring a single quadrature, determining its statistics to be Gaussian, and inferring its boundary conditions, mutual information can be measured, provided the system linearly evolves within resolvable time scales. The dispersion relation can be reconstructed, and linear evolution can be verified at no additional experimental effort. A low-pass filter may additionally be applied to remove non-linear behaviour at high frequencies.


The state of system may also be manipulated in many advanced quantum simulators. Whilst state preparation and control has been well established through Floquet engineering in Bose-Einstein condensates~\cite{Goldman_2015, Weitenberg2021, Ji_2022}, nano-fabrication and magnetic-field shaping techniques have shown promising demonstrations of state-selective control in superfluid helium systems~\cite{Spence_2021, He2020, Forstner_2019}. Progress in state preparation holds promise for studying information both in a plethora of states or subject to a quench through the rapid variation of the height or propagation speed.

It is worth noting that classical information methods may be employed when the states are dominated by classical effects~\cite{Haas2024, Deller2024a, Deller2024b}. However, beyond these classically dominated states, we wish to examine the feasibility of experimental measurements in the quantum regime, defined by modes satisfying $\hbar\omega_m\gtrsim k_{\text{B}}T$. For film thicknesses of approximately $50$nm within a micrometric confinement, the quantum regime corresponds to temperatures $T\lesssim 1 \mu$K for the lowest-frequency modes in the system. Measurements of height perturbations at currently feasible temperatures may soon allow us to measure information in states where high-momentum modes lie within the quantum regime, whilst the low-momentum modes remain classically dominated. This offers a platform for studying information at the interface between classical and quantum physics. By considering these regimes separately, we can distinguish the classical and quantum contributions to the scaling of mutual information.

Realising an experimental probe for mutual information in thin-film superfluid helium fully within the quantum regime requires both the stable control of ultra-low temperatures, where third-sound modes can be described by discrete numbers of phonon excitations, and a minimally invasive measurement scheme with spatial and temporal resolution sufficient to operate on the film in the zero-temperature limit. Solutions to each of these have been developed or are in development for use in superfluid helium experiments. The zero-temperature limit in superfluid helium-$4$, where the normal-component fraction becomes negligible, is of order $10$--$100$mK \cite{brooks1977calculated, donnelly1998observed}, well within reach of a modern dilution refrigerator~\cite{zu2022development}. At these temperatures, phonon occupation numbers in third sound modes are expected to be in the hundreds~\cite{Atkins1959}. To reach lower temperatures, where quantised phononic excitations become relevant, a nuclear demagnetisation stage is needed. Nuclear demagnetisation stages in dry dilution refrigerators have recently been developed~\cite{nyeki2022high}, allowing such experiments to be performed with optical access at hundreds of $\mu$K. The non-invasive imaging of third-sound modes at a few micro-Kelvins requires low optical powers, as in recent implementations of digital holography on trapped atomic clouds at ultra-low temperatures~\cite{Blaznik_2024, Smits_2020}. This combination of an efficient measurement within an ultra-low temperature environment provides an experimental pathway to measuring mutual information in thin films of superfluid helium, even deep within the quantum realm.

\section{\label{subsection::Conclusions} Conclusion}
\vspace{-4mm}
We have examined the accessibility of information in experimentally feasible quantum field simulators. We provide a methodology for the analysis of area laws, identifying a pathway towards realising the characterisation of information in thin films of superfluid helium, pushing towards the quantum-dominated regime. Our approach enables the extraction of covariance matrices for Gaussian states, providing a foundation for experimentally probing environment interactions~\cite{Menonifmmode_2024, Aimet_2024_Landauer} and quantum thermalisation~\cite{Trotzky_2012, Geiger2014, Kaufman_2016, Islam_2015} through relative entropies~\cite{Cover_thomas_2006, Vedral_2002, Schrofl_2024, Genoni_2008, Floerchinger2023} and quantum discord~\cite{Adesso_2010}.

\section*{ACKNOWLEDGMENTS}
\vspace{-4mm}
We want to thank Jorma Louko, William G. Unruh, Tobias Haas, Sebastian Erne, Cisco Gooding, and Daniel Carney for fruitful discussions and feedback. MTJ, MT, and SW extend their appreciation to Science and Technology Facilities Council on Quantum Simulators for Fundamental Physics (ST/T006900/1) as part of the UKRI Quantum Technologies for Fundamental Physics programme. MTJ, VSB, CG, and SW gratefully acknowledge the support of the Leverhulme Research Leadership Award (RL2019-020). The work of CRDB was supported by United Kingdom Research and Innovation Engineering and Physical Sciences Research Council (EPSRC) [grant number EP/W524402/1]. SW acknowledges the Royal Society University Research Fellowship (UF120112,RF/ERE/210198, RGF/EA/180286, RGF/EA/181015) and partial support by the Science and Technology Facilities Council (Theory Consolidated Grant ST/P000703/1).

\appendix
\onecolumngrid

\section{Two-point function reconstruction equation}\label{app:reconstruction}
We calculate the time evolution of the two-point functions of the field and its conjugate momentum. Substituting~\eqref{eq:general decomp} and~\eqref{eq:general mode t evoln} into the expression for the two-point function, we find
\begin{equation}\label{eq:general reconstruction phiphi}
    \begin{split}
        \langle\hat{\phi}(t,\bm{x}_i)\hat{\phi}(t,\bm{x}_j)\rangle ~=~ \sum_{mn} \frac{c}{K\sqrt{\omega_m \omega_n}} \Bigl[ &g_m(\bm{x}_i)g_n(\bm{x}_j)\cos\lr{\omega_m t}\cos\lr{\omega_n t}\tilde{Q}_{mn}(0) \\
        +& g_m(\bm{x}_i)g_n(\bm{x}_j)\sin\lr{\omega_m t}\sin\lr{\omega_n t}\tilde{P}_{mn}(0)\\
        -& \lr{g_m(\bm{x}_i)g_n(\bm{x}_j) + g_n(\bm{x}_i)g_m(\bm{x}_j)}\cos\lr{\omega_m t}\sin\lr{\omega_n t} \tilde{R}_{mn}(0) \Bigr] \,,
    \end{split}
\end{equation}
\begin{equation}\label{eq:general reconstruction etaeta}
    \begin{split}
        \langle\hat{\eta}(t,\bm{x}_i)\hat{\eta}(t,\bm{x}_j)\rangle ~=~ \sum_{mn} \frac{K\sqrt{\omega_m \omega_n}}{c} \Bigl[ &g_m(\bm{x}_i)g_n(\bm{x}_j)\cos\lr{\omega_m t}\cos\lr{\omega_n t}\tilde{P}_{mn}(0)\\
        +& g_m(\bm{x}_i)g_n(\bm{x}_j)\sin\lr{\omega_m t}\sin\lr{\omega_n t}\tilde{Q}_{mn}(0)\\
        -& \lr{g_m(\bm{x}_i)g_n(\bm{x}_j) + g_n(\bm{x}_i)g_m(\bm{x}_j)}\cos\lr{\omega_m t}\sin\lr{\omega_n t} \tilde{R}_{mn}(0) \Bigr] \,,
    \end{split}
\end{equation}where we have identified
\begin{equation}
\begin{split}
    \tilde{Q}_{mn}(0)~&=~\langle\hat{\phi}_m(0)\hat{\phi}_n(0)\rangle\,,\\
    \tilde{P}_{mn}(0)~&=~\langle\hat{\eta}_m(0)\hat{\eta}_n(0)\rangle\,,\\
    \tilde{R}_{mn}(0)~&=~\frac12\langle\{\hat{\phi}_m(0)\hat{\eta}_n(0)\}\rangle\,.
\end{split}
\end{equation}

\section{Robin boundaries}\label{app:Robin boundaries}
The theory described by the Hamiltonian~\eqref{eq:general H} is only compatible with fields obeying Dirichlet ($\hat{\phi}|_{\partial V}=0$) or Neumann ($\bm{n}\cdot\nabla\hat{\phi}|_{\partial V}=0$) boundary conditions at $\partial V$. Both Dirichlet and Neumann boundary conditions are special cases of the more general Robin family of boundary conditions, characterised by a single parameter $\alpha$ with units of inverse length,
\begin{equation}\label{eq: Robin boundary}
    \alpha\hat{\phi}|_{\partial V}~=~\bm{n}\cdot\nabla\hat{\phi}|_{\partial V}\,.
\end{equation}Dirichlet and Neumann boundary conditions are recovered as $\alpha\to\infty$ and $\alpha\to0$ respectively. The Hamiltonian density describing a theory compatible with Robin boundary conditions is given by
\begin{equation}\label{eq:H gen Robin}
    \hat{\mathcal{H}}~=~\frac{\hbar c}{2}\left[K|\nabla\hat{\phi}|^2+\frac1K\hat{\eta}^2+\frac{c^2}{\hbar^2}M^2K\hat{\phi}^2-\alpha K \delta(f(\bm{x}))\hat{\phi}^2\right]\,,
\end{equation}where $f(\bm{x})=0$ defines the boundary.


The additional term in~\eqref{eq:H gen Robin} may be interpreted as an effective mass localised on the boundary~\cite{deAlbuquerque:2003qbk}. However, care must be taken with this interpretation as the equations of motion are still given by the Klein-Gordon equation~\eqref{eq:general EoMs}. In particular, the mass of the field is given by $M$.


We decompose the field and momentum as in~\eqref{eq:general decomp}. Integrating over the spatial volume $V$, we recover the diagonalised quantum harmonic oscillator Hamiltonian~\eqref{eq:general QHO}. Again, the oscillator frequency is given by~\eqref{eq:general disperion}, which may be anticipated on physical grounds; the boundary conditions may not change the dispersion relation. What distinguishes the different boundary conditions, however, are the wavenumbers $k_m$ in the system. Formally, these wavenumbers are functions of the Robin parameter $\alpha$ that smoothly interpolates between the Dirichlet and Neumann cases.

\section{Mutual information maps}\label{app:MI maps}
\begin{figure*}[h!]
    \centering
    \resizebox{\linewidth}{!}{\includegraphics[trim=0 70 0 90, clip]{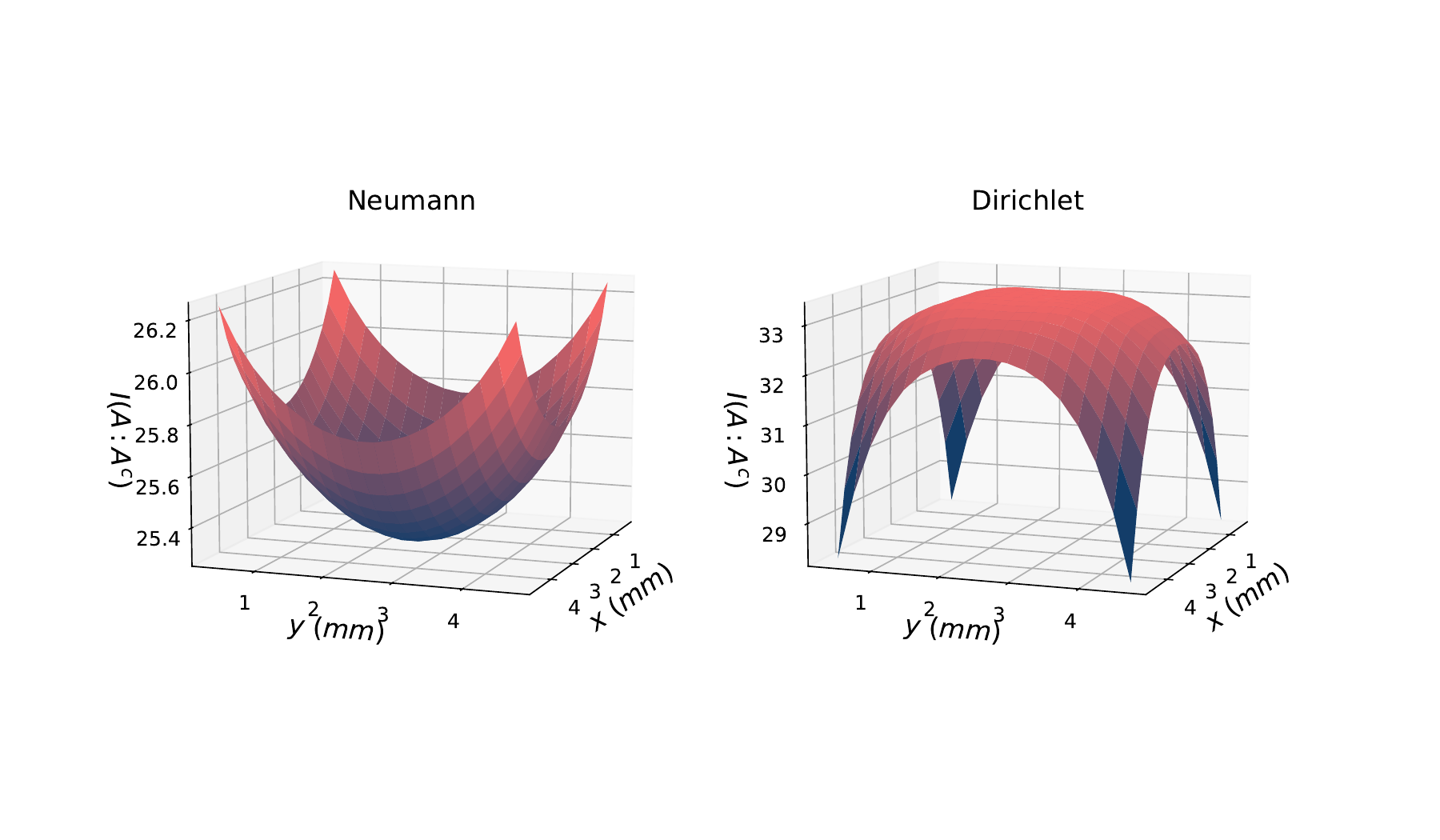}}
    \caption{\justifying Maps of local mutual information in a thermal state. Parameters: $T=0.3$K, $c_3=0.12$ms$^{-1}$, $K=2.21\time 10^{14}$m, $L_x=L_y=5$mm, $N_x=N_y=20$, and $\Delta x=\Delta y=0.25$mm.
    }
    \label{fig:MI maps}
\end{figure*}

We consider the local information in systems with Neumann or Dirichlet boundary conditions. We define subsystem $A$ as a single spatial point, with subsystem $B=A^c$. In order to maintain a constant boundary area and number of edges within the definitions of subsystems, we exclude the boundary points from the analysis. The results are shown in Fig.~\ref{fig:MI maps}.

The boundary conditions determine the distribution of local mutual information, with qualitatively different results for the Dirichlet and Neumann cases. Whilst we have not computed the local mutual information for Robin boundary conditions, we expect this to smoothly interpolate between Dirichlet and Neumann. We note that choosing subsystem $A$ to be a single point produces MI equivalent to the maximal amount obtainable from an optimal local measurement.

\twocolumngrid

\bibliography{bibphy}
\end{document}